\documentclass[sigconf]{acmart}

\AtBeginDocument{%
  }

\setcopyright{acmlicensed}
\copyrightyear{2026}
\acmYear{2026}
\acmDOI{XXXXXXX.XXXXXXX}
\acmConference[FSE SEET '26]{ACM SIGSOFT International Symposium on Foundations of Software Engineering (FSE) - Software Engineering Education Track (SEET)}{Jul 2026}{Montreal, QC, Canada}
\acmBooktitle{Companion Proceedings of the 34th ACM Symposium on the Foundations of Software Engineering (FSE '26), July 5 - 9, 2026 Montreal, Canada}

\acmISBN{XXXX-XX-XXXX-XXXX-X/26/07}

\usepackage{balance}
\usepackage{graphicx}
\usepackage{amsmath,amsfonts}
\usepackage{algorithmic}
\usepackage{algorithm}
\usepackage{array}
\usepackage{textcomp}
\usepackage{stfloats}
\usepackage{url}
\usepackage{verbatim}
\usepackage{graphicx}
\usepackage{multirow}
\usepackage{listings}
\usepackage{subcaption}
\usepackage[T1]{fontenc}
\usepackage{multirow}
\usepackage{array}
\usepackage{threeparttable}
\usepackage{tabularray}
\usepackage{hyperref}
\usepackage{tabularx}
\usepackage{array}
\usepackage{booktabs}

\begin{document}

\title{Large Language Models for Software Testing Education: an Experience Report}

\author{Peng Yang}
\affiliation{%
  \institution{South China Normal University \\ Guangzhou Polytechnic University}
  \city{Guangzhou}
  \country{China}}
\email{yangpeng@m.scnu.edu.cn}

\author{Yunfeng Zhu}
\affiliation{%
  \institution{State Key Laboratory of Novel Software Technology, Nanjing University}
  \city{Nanjing}
  \country{China}
}
\email{yunfengzhu@smail.nju.edu.cn}

\author{Chao Chang}
\affiliation{%
  \institution{South China Normal University}
  \city{Guangzhou}
  \country{China}
}
\email{changchao@m.scnu.edu.cn}

\author{Shengcheng Yu}
\affiliation{%
  \institution{State Key Laboratory of Novel Software Technology, Nanjing University}
  \city{Nanjing}
  \country{China}
}
\email{yusc@nju.edu.cn}

\author{Zhenyu Chen}
\affiliation{%
  \institution{Mooctest Inc.}
  \city{Nanjing}
  \country{China}
}
\email{zychen@nju.edu.cn}

\author{Yong Tang}
\affiliation{%
  \institution{South China Normal University}
  \city{Guangzhou}
  \country{China}
}
\email{ytang@m.scnu.edu.cn}

\begin{abstract}
The rapid integration of Large Language Models (LLMs) into software engineering practice is reshaping how software testing activities are performed. LLMs are increasingly used to support software testing. Consequently, software testing education must evolve to prepare students for this new paradigm. However, while students have already begun to use LLMs in an ad hoc manner for testing tasks, there is limited empirical understanding of how such usage influences their testing behaviors, judgment, and learning outcomes. It is necessary to conduct a systematic investigation into how students learn to evaluate, control, and refine LLM-assisted testing results.

This paper presents a mixed-methods, two-phase exploratory study on human–LLM collaboration in software testing education. In Phase~I, we analyze classroom learning artifacts and interaction records from 15 students, together with a large-scale survey conducted in a national software testing competition (337 valid responses), to identify recurring prompt-related difficulties across testing tasks. The results reveal systematic interaction breakdowns, including missing contextual information, insufficient constraints, rigid one-shot prompting, and limited strategy-driven iteration, with automated test script generation emerging as a particularly heterogeneous and effort-intensive interaction context. Building on these findings, Phase~II conducts an illustrative classroom practice that operationalizes the observed breakdowns into a lightweight, stage-aware prompt scaffold for test script generation, guiding students to explicitly articulate execution-relevant information such as environmental assumptions, interaction grounding, synchronization, and validation intent, and reporting descriptive shifts in students’ testing-related articulation when interacting with LLMs.

This paper empirically characterizes task-dependent difficulties and interaction patterns in students’ use of LLMs for software testing from a learning behavior perspective. It also provides descriptive evidence distinguishing relatively stable interaction contexts from more heterogeneous and unstable ones across testing tasks. Moreover, this paper illustrates a pedagogical direction for translating observed LLM usage difficulties into instructional support, grounded in classroom practice rather than tool-centric optimization. In summary, these findings offer empirical grounding for integrating LLMs into software testing education in a learning-oriented and pedagogically informed manner.

\end{abstract}

\ccsdesc[500]{Software and its engineering~Software testing and debugging}

\keywords{Large Language Models; Software Testing Education; Task Decomposition; Prompt Engineering; Script Generation}

\maketitle

\section{Introduction}
With the rapid adoption of large language models (LLMs) in software engineering practice~\cite{daun2023chatgpt}, proficiency in using LLMs has become an increasingly important competence for software developers. Across software testing activities—including requirements understanding, test design, test case generation, and automated test script development—LLMs have demonstrated the potential to lower operational barriers and improve efficiency. As these capabilities are progressively integrated into industrial workflows, future software engineers are expected not only to access LLM-based tools but also to use them effectively and appropriately as part of everyday development and testing practices~\cite{zhang2023survey}. From an industry perspective, the ability to collaborate with LLMs is therefore emerging as a practical skill that software engineering education can no longer ignore.

From an educational perspective, however, LLMs are not merely neutral productivity tools~\cite{vierhauser2024towards}. Students have already begun to incorporate LLMs into software testing–related learning activities~\cite{ wieser2023investigating, sengul2024software}, embedding them directly into task execution workflows in authentic course settings. Prior research has examined the impact of LLMs and other generative AI tools by comparing learning outcomes or task performance between AI-assisted and traditional instructional supports~\cite{kazemitabaar2023studying, tang2024chatgpt}. While these studies provide valuable evidence regarding effectiveness, they often treat LLMs as monolithic tools and primarily focus on whether they are useful~\cite{denny2023can}. Consequently, much less is known about how students actually use LLMs in concrete testing tasks, what difficulties they encounter, and whether such usage genuinely supports learning processes such as understanding, internalization, and the development of independent testing judgment. Understanding students’ real-world interaction patterns with LLMs thus represents a critical prerequisite for designing pedagogically sound approaches to integrating LLMs into software testing education.

Software testing learning is inherently task-heterogeneous~\cite{kokotsaki2016project, majumdar2024mining}. Core testing activities—including test scope definition and strategy formulation, test case design, and automated test script implementation—differ substantially in the types of knowledge involved, the levels of abstraction required, and the degree of operational precision demanded. Emerging studies suggest that students’ reliance on large language models (LLMs), their interaction styles, and their perceived benefits may vary across such tasks~\cite{macneil2024synthetic}. However, it remains unclear whether these task-dependent differences exhibit systematic or stable characteristics or how they relate to students’ learning experiences and the development of testing-related competencies.

Against this backdrop, we shift the focus from evaluating the technical performance of LLMs in software testing to examining students’ learning-oriented use of LLMs in software testing education. Rather than asking whether LLMs can generate correct or high-quality testing artifacts, we investigate how students incorporate LLMs into their problem-solving processes. In particular, we examine whether LLMs function as cognitive scaffolds that support task understanding, reasoning about testing logic, and knowledge construction, or whether they are primarily used as substitutes for independent reasoning and code generation~\cite{denny2023can}. From an educational perspective, this distinction is critical for understanding when and how LLM use may meaningfully contribute to learning rather than merely task completion.

Despite growing interest in LLM-assisted software engineering education~\cite{selwyn2019should}, systematic empirical evidence characterizing students’ task-dependent LLM usage behaviors in software testing contexts remains limited~\cite{vierhauser2024towards, macneil2024synthetic}. To address this gap, we conduct an exploratory mixed-methods study guided by the following research questions:
\begin{itemize}
\item \textbf{RQ1:} What major difficulties do students encounter when using LLMs across different software testing tasks, and do these difficulties exhibit stable, task-dependent patterns?
\item \textbf{RQ2:} How prevalent are these usage patterns and difficulties among students, and how does students’ perceived effectiveness of LLMs differ across testing task types?
\item \textbf{RQ3:} How can the difficulties identified in Phase~I be translated into instructional support for LLM-assisted test script generation, and what descriptive changes are observed in students’ articulation of execution-relevant information when such support is introduced?
\end{itemize}

This paper makes the following contributions to software engineering education. First, from a learning behavior perspective, we empirically characterize the difficulties and interaction patterns that emerge when students use LLMs across heterogeneous software testing tasks. Second, by triangulating classroom-based evidence with large-scale survey data, we provide comparative empirical insights into the differing degrees of uncertainty associated with text-oriented testing tasks and automated test script generation. Third, through an illustrative instructional exploration, we demonstrate that the observed usage difficulties of LLMs are pedagogically actionable, offering practical insights for the responsible and learning-oriented integration of LLMs into software testing education. Supplementary materials related to this study—including detailed protocols, example prompts, and the appendix—are publicly available at: \url{https://anonymous.4open.science/r/B79E/README.md}.

\section{Background and Related Work}

\subsection{Software Testing Education}
Software testing education has continuously evolved through the adoption of diverse pedagogical approaches and instructional tools~\cite{wang2024software}. Traditional teaching practices have been extended with automated assessment tools such as PROGTEST~\cite{souza2014assess, amalfitano2023artificial}, mutation testing–based instructional techniques~\cite{towey2015teaching}, and game-based learning environments such as Code Defenders~\cite{clegg2017mut}. In addition, problem-based learning (PBL)~\cite{cheiran2017problem, delgado2022interevo} and adaptive blended learning approaches~\cite{elgrably2022quasi} have been introduced to provide students with more personalized and practice-oriented learning experiences.

Project-based learning has emerged as a central instructional strategy for bridging theoretical knowledge and practical testing skills by situating students in realistic software testing scenarios~\cite{koutcheme2024using, kokotsaki2016project, eisty2025testing}. Prior empirical studies suggest that such approaches can enhance students’ conceptual understanding and better prepare them for professional testing environments~\cite{majumdar2024mining, michaeli2017test}. Despite these advances, comparatively little is known about how students’ interaction practices evolve when emerging AI-based tools are embedded into different testing tasks, particularly in authentic course settings where multiple task types coexist.

\subsection{LLMs in Software Engineering Education}

More broadly, LLMs have been characterized as foundation models—pretrained on large-scale data and adaptable to a wide range of downstream tasks—bringing both new opportunities and inherited risks to educational contexts~\cite{bommasani2021opportunities}. Their integration into software engineering education has prompted increasing attention to how students learn, practice, and receive feedback. Prior work has explored the use of LLMs for generating learning materials, providing personalized recommendations, and supporting instructional scalability~\cite{vierhauser2024towards}, while also cautioning against potential long-term impacts on learning and skill development~\cite{denny2023can, prather2023robots}.

Empirical studies have examined how LLM-based systems can assist students in programming tasks, problem solving, and formative feedback generation~\cite{wieser2023investigating, DBLP:journals/te/NeumannYSDJ25}, as well as in tasks such as algorithm reasoning and pseudocode verification~\cite{balse2023exploring}. Other work highlights LLMs’ potential to interpret code-related issues and provide contextualized guidance~\cite{koutcheme2024using}. However, much of the existing literature emphasizes tool capabilities or learning outcomes, with relatively limited attention to how students actually interact with LLMs during task execution, or how interaction breakdowns may emerge in different learning contexts.

\subsection{LLMs in Software Testing Education}
More recently, researchers have begun to explore the application of LLMs specifically within software testing education~\cite{macneil2024synthetic, yu2023llm, wang2024software}. Prior studies suggest that LLMs can support a range of testing-related activities, including requirements understanding, identification of potential issues~\cite{xue2024llm4fin}, and test case generation~\cite{Sindre_2023}. LLMs have also been explored as supports for specifying user interactions and generating test steps or scripts, potentially exposing students to contemporary testing practices.

At the same time, integrating LLMs into software testing education introduces challenges that differ from those observed in general programming contexts~\cite{balse2023exploring, jalil2023chatgpt}. Effective use of LLMs requires students to formulate prompts that reflect both software testing principles and an understanding of LLM interaction characteristics~\cite{daun2023chatgpt}. In addition, aligning LLM-generated artifacts with existing testing tools and frameworks can impose structural and process-level overheads~\cite{wieser2023investigating}. Despite growing interest, systematic empirical evidence remains limited regarding where task-dependent interaction breakdowns occur during LLM-assisted testing and what cognitive or operational costs students incur across different testing tasks.

\section{Method}

\subsection{Overview of Research Design}

This study adopts a mixed-methods research design to investigate how students use LLMs in course-related software testing tasks and to identify challenges that emerge during this process. Unlike studies that focus on LLM performance or output quality, our focus is on students’ learning-oriented usage behaviors, specifically where difficulties arise, how they manifest across different task types, and whether such difficulties exhibit systematic, task-dependent patterns. Understanding these patterns is critical for informing instructional design that can guide students in integrating LLMs effectively into software testing workflows.

\begin{figure*}[htbp]
    \centering
    \includegraphics[width=0.7\linewidth]{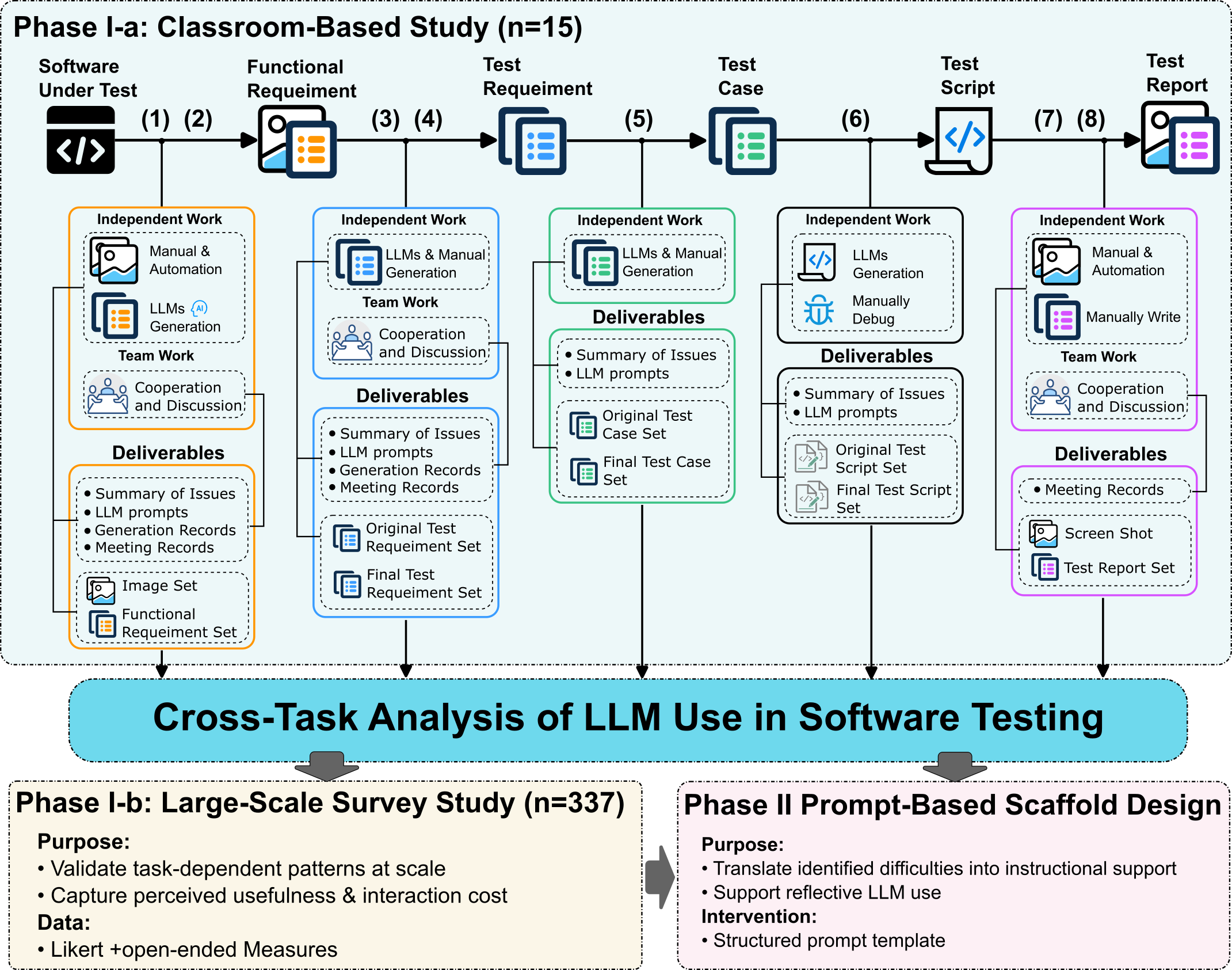}
    \caption{Overview of the study design. Phase I combines classroom study and survey analysis to identify task-dependent LLM usage patterns and difficulties. Phase II illustrates a stage-aware instructional scaffold informed by Phase I, highlighting how LLM-supported tasks can be made pedagogically actionable.}
    \label{fig:main}
\end{figure*}

The study consists of two phases (Phase~I and Phase~II), as illustrated in \textbf{Figure \ref{fig:main}}.
\begin{itemize}
    \item \textbf{Phase I} is exploratory, combining a classroom-based study with a large-scale survey to characterize students’ LLM usage behaviors and task-dependent differences from complementary depth and breadth perspectives. This phase emphasizes empirical observation and descriptive analysis, identifying recurring interaction patterns and task-specific challenges.
    \item \textbf{Phase II} builds on Phase I findings through an illustrative instructional practice, exploring whether observed difficulties can be addressed pedagogically using a lightweight prompt scaffold. This phase highlights the potential teachability of LLM usage challenges and examines how guided articulation of task-relevant information can shift students’ interactions from ad-hoc delegation to reflective, structured problem solving.
\end{itemize}

Phase I emphasizes phenomenon characterization, while Phase II demonstrates the instructional implications of these observations, without making claims about causal learning gains.

\subsection{Study Context and Participants}

The classroom study was conducted in an advanced software testing course for undergraduate software engineering students. The course curriculum included test strategy formulation, test case design, and automated test script development, with hands-on projects serving as the primary learning activities. Students were allowed to use LLMs as auxiliary tools at their discretion, without formal guidance on prompt engineering or structured interaction strategies. This design aimed to capture naturalistic, task-driven use of LLMs in authentic learning contexts, reflecting real-world variability in student behaviors.

A total of 15 third-year students voluntarily participated. All had completed foundational software engineering and testing courses but had not received prior instruction in prompt engineering or AI-assisted testing. This setting enabled observation of relatively unconstrained LLM usage, providing insight into common breakdowns and strategies students adopt when working independently.

To enhance external validity, we conducted a large-scale competition survey with 337 valid responses from students across multiple universities. Participants reported their experience with software testing and LLM usage, including frequency, perceived effectiveness, interaction costs, and learning outcomes. This dataset complements the classroom study, allowing cross-context examination of whether task-dependent patterns observed in the classroom generalize to a broader population with diverse prior experience and institutional settings.

\subsection{Phase I:Understanding Students’ LLM Usage}

Phase I aimed to systematically characterize how students engage with LLMs across software testing tasks and to identify recurring difficulties. By integrating process-level classroom data with competition survey data, this phase achieves methodological complementarity, combining in-depth qualitative insights with broader quantitative trends.

Data sources included student-generated artifacts, LLM interaction logs, classroom observations, and survey responses, enabling triangulation across behaviors, perceptions, and outputs. This design provides a robust foundation for identifying recurring challenges and task-dependent usage patterns.

\subsubsection{Task Taxonomy and Cross-Source Mapping}
\label{sec:task-mapping}

Because the classroom study and the competition survey differ in granularity, we define a unified taxonomy to keep cross-source interpretations consistent. We group activities into two task families according to (i) the primary representation of produced artifacts and (ii) the specificity required for obtaining usable outputs from LLM prompting.

\paragraph{Text-oriented tasks.}
These tasks primarily produce natural language testing artifacts and involve reasoning and documentation (e.g., requirements-related analysis, test planning/design, test case description, and reporting). In the classroom study, the task comparisons reported in Figure~\ref{fig:Bubble} focus on requirement analysis and test case generation. In the competition survey, we use Q4\_1--Q4\_3 (requirements analysis, test planning/design, and reporting) as corresponding indicators for this family.

\paragraph{Script-oriented tasks.}
These tasks produce executable automated test scripts (e.g., based on web/mobile automation frameworks) and typically require explicit execution specifications (environment/framework versions), interaction grounding (element locators), synchronization logic, and assertions to obtain runnable results. In the classroom study, this corresponds to automated test script generation; in the competition survey, this corresponds to Q4\_4 (test script generation).

\paragraph{Alignment scope.}
Throughout the paper, we interpret cross-source differences at the \emph{task-family level} (text-oriented vs.\ script-oriented). We do not assume a one-to-one equivalence between each classroom task and each survey item; instead, survey items serve as complementary, family-level indicators that contextualize the classroom observations.

\subsubsection{Classroom-Based Exploratory Study}

In the classroom study, we collected multiple data sources, including student-submitted testing artifacts (e.g., test plans, test cases, and automated test scripts), interaction records with LLMs (i.e., prompts and iterative prompt revisions), as well as execution results and error manifestations of generated scripts.

Through iterative analysis of students’ prompts and interaction logs, we employed a combination of open coding and axial coding to categorize recurring issues that emerged during LLM usage. The analysis aim to identify \textbf{which types of missing information, expression patterns, or interaction strategies repeatedly led to difficulties across different tasks}.

This analysis resulted in six categories of prompt-related issues that consistently appeared in the study context. We further compared the distribution of these issues across different testing tasks to explore their task-dependent characteristics.The coding scheme was iteratively refined through repeated examination of student artifacts and discussions among the research team until a stable set of issue categories emerged.

\subsubsection{Competition Survey Study}

To complement the classroom findings, we designed and administered a competition survey. The survey focused on students’ frequency of LLM usage across different testing tasks, perceived effectiveness, interaction cost, and attitudes toward the learning impact of LLM use.

The survey combined Likert-scale items with open-ended questions. Quantitative items were used to compare perceived differences in LLM usage across tasks, while open-ended responses captured students’ subjective judgments regarding failure causes, missing critical information, and potential learning risks.

In the analysis, we focused on \textbf{relative differences and consistent trends across task types}, rather than precise point estimates for individual items. This strategy positioned the survey data as complementary evidence supporting classroom observations, rather than as an independent source of confirmatory conclusions, especially for item-level task comparisons.

\subsection{Phase II: Illustrative Instructional Scaffold}

Building on the task-dependent difficulties identified in Phase~I, Phase~II explores whether these challenges—many of which manifest at specific stages of the software testing process—can be pedagogically addressed through structured instructional guidance. This phase is intended to illustrate the teachability of observed difficulties and to examine how stage-aware interventions can help students articulate testing assumptions and reasoning when interacting with LLMs.

In particular, recurring patterns identified in automated test script generation—a task type characterized by high interaction cost, variable success, and complex execution dependencies—directly informed the dimensions emphasized in the prompt-based scaffold implemented in this phase.

\subsubsection{Rationale for Instructional Support}

Phase~I findings indicated that script-oriented tasks consistently exhibited higher uncertainty, longer interaction rounds, and greater failure rates across classroom and survey data. Qualitative analysis suggested that these challenges were frequently associated with insufficient articulation of task-relevant information, including:
\begin{itemize}
\item Testing context (environment assumptions and constraints)
\item Execution assumptions (framework, platform, synchronization)
\item Validation intent (expected outcomes and success criteria)
\end{itemize}

Rather than testing formal hypotheses about instructional effectiveness, Phase~II investigates whether providing structured prompts that explicitly guide students to organize and externalize key testing information prior to LLM interaction can alter recurring problem patterns. This approach prioritizes learning-oriented reflection, allowing students to reason systematically about task requirements and LLM outputs.

\subsubsection{Illustration of a Prompt-Based Scaffold}

In the instructional practice, we introduced a lightweight structured prompt scaffold in the form of a template. The scaffold guided students to explicitly specify testing information aligned with different stages of the testing process, including environment assumptions (context specification), element locators (interaction grounding), operation flows and synchronization logic (execution planning), and verification logic (validation) before generating test scripts.

Importantly, the scaffold did not prescribe specific testing strategies or implementation approaches, nor did it restrict how students interacted with LLMs. Instead, it functioned as a \textbf{reflective tool} to help students externalize testing assumptions and decision rationales during LLM interaction.

\subsubsection{Scope and illustration Approach}

The analysis in Phase~II focuses on changes in error types and manifestations associated with different testing stages before and after the introduction of instructional guidance, rather than on success rates, performance metrics, or quantified learning gains. By comparing recurring error patterns in generated scripts, we examined which issues became less frequent following the introduction of the scaffold. This analysis focuses on how students’ engagement with LLMs shifted across testing stages, rather than on overall script correctness or performance.

Phase~II serves an \textbf{illustrative rather than evaluative} role. Its purpose is to demonstrate that some difficulties identified in Phase~I may be pedagogically addressable rather than solely determined by model capability, thereby providing an empirical basis for future, more systematic instructional research.

\section{Results}

\subsection{Phase I: Understanding Students’ LLM Usage in Software Testing}

Phase~I aims to characterize how students use LLMs across different software testing tasks and to identify task-related difficulty patterns and usage practices. The results integrate analyses of learning artifacts from the classroom study with data from a competition survey, providing complementary evidence from both fine-grained, individual-level observations and population-level self-reported trends.

\subsubsection{Task-Dependent Difficulties in LLM-Assisted Testing}

Through iterative coding of students’ prompts and interaction logs collected in the classroom study, we identified six recurring categories of prompt-related issues: rigid usage patterns, missing context, lack of interaction, insufficient constraints, vague task descriptions, and missing input/output specifications.

\begin{figure}[htbp]
    \centering
    \includegraphics[width=0.8\linewidth]{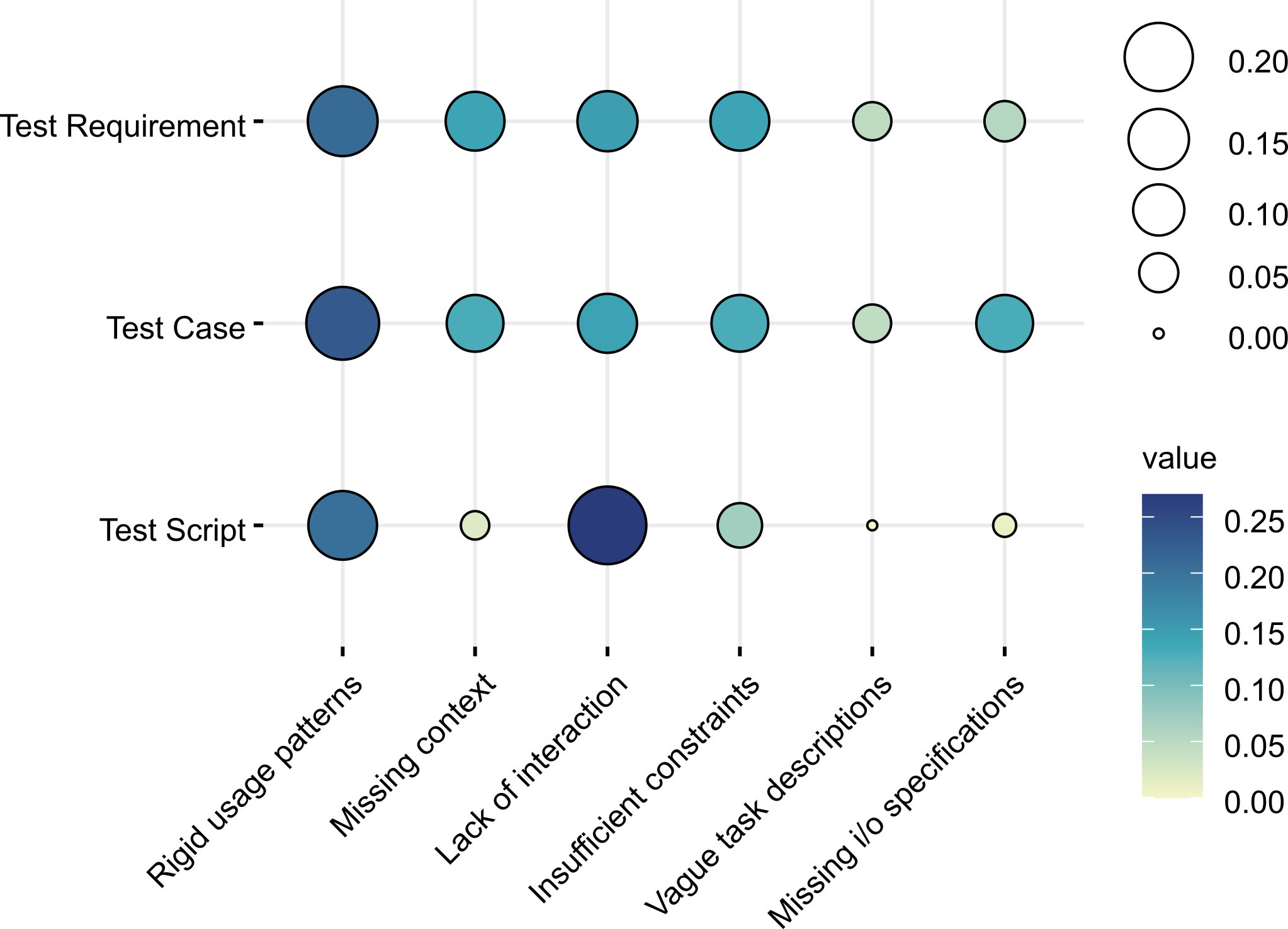}
    \caption{Distribution of prompt design issue categories across different software testing tasks}
    \label{fig:Bubble}
\end{figure}

Figure~\ref{fig:Bubble} visualizes the relative distribution of these prompt-related issues across three representative testing tasks: test requirement analysis, test case generation, and automated test script generation. Bubble size encodes the proportion of each issue category within a given task, allowing direct comparison of issue salience across tasks.

The identification and categorization of these prompt-related issues were based on a systematic qualitative coding process applied to classroom artifacts and interaction logs. Details of the coding procedure and category definitions are provided in supplementary material.

Across all tasks, \emph{rigid usage patterns} constitute one of the most prevalent issues, accounting for roughly one-fifth of observed problems in each task category. This consistency suggests a shared baseline difficulty: many students tended to treat LLMs as one-shot answer generators rather than as interactive collaborators, regardless of task type.

Beyond this common pattern, clear task-dependent differences emerge. For text-oriented tasks—namely test requirement analysis and test case generation—the overall distributions are highly similar. In both tasks, issues related to missing context, insufficient constraints, and lack of interaction appear at comparable levels, each occupying a moderate proportion of observed problems. This relatively balanced distribution indicates that students approached these tasks using familiar natural-language reasoning strategies, resulting in similar prompt construction behaviors across tasks.

In contrast, automated test script generation exhibits a markedly different structural profile. Most notably, \emph{lack of interaction} becomes the dominant issue, accounting for approximately one-quarter of all observed problems—nearly double the proportion observed in text-oriented tasks. This sharp increase suggests that students were substantially less likely to engage in iterative refinement when generating scripts, often expecting complete and executable solutions from a single prompt.

Conversely, \emph{missing context} issues are substantially less frequent in script generation, appearing only sporadically compared to their presence in text-oriented tasks. This pattern indicates a qualitative shift rather than an improvement: instead of omitting context unintentionally, students appeared to assume that critical execution details—such as environment configuration or element locators—could be implicitly inferred by the model. As a result, contextual gaps were less visible at the prompt level but later surfaced as execution-time failures.

Issues related to insufficient constraints and missing input/output specifications remain present in script generation, but their relative prominence is overshadowed by interaction-related deficiencies. Taken together, these distributions suggest that prompt-related difficulties in script generation are not simply more frequent, but structurally different in nature.

These task-dependent distributions indicate that prompt quality issues are not generic deficiencies but reflect deeper differences in how students conceptualize LLM interaction across task types. Script-oriented tasks place higher demands on task decomposition, iterative clarification, and explicit constraint articulation. When these demands are unmet, prompt-level ambiguities propagate into downstream failures, helping to explain why script generation later emerges as a high-uncertainty and high-cost activity in both perceived effectiveness and interaction outcomes.

Taken together, these task-dependent difficulty patterns raise a further question: whether the structural differences observed in students’ LLM interactions are also reflected in how students \emph{perceive} the usefulness and reliability of LLM assistance across different testing tasks. To address this question, we next examine students’ perceived effectiveness of LLMs through survey data.

The relative proportions visualized in Figure~\ref{fig:Bubble} are derived from aggregated coding results across tasks. The construction rationale and encoding of the bubble chart are described in supplementary material. 

\subsubsection{Perceived Effectiveness as a Signal of Task Instability}

Beyond prompt-level difficulties observed in classroom artifacts, we examined how students \emph{perceived} the usefulness and reliability of LLM assistance across different software testing tasks.
These perceptions were collected through a task-specific Likert-scale questionnaire administered in the competition survey and are interpreted here as reflections of students’ subjective confidence and perceived helpfulness during task execution, rather than as indicators of LLM performance.
The corresponding Likert-scale survey items and their exact wording are provided in supplementary material.

\begin{figure}[htbp]
    \centering
    \includegraphics[width=1\linewidth]{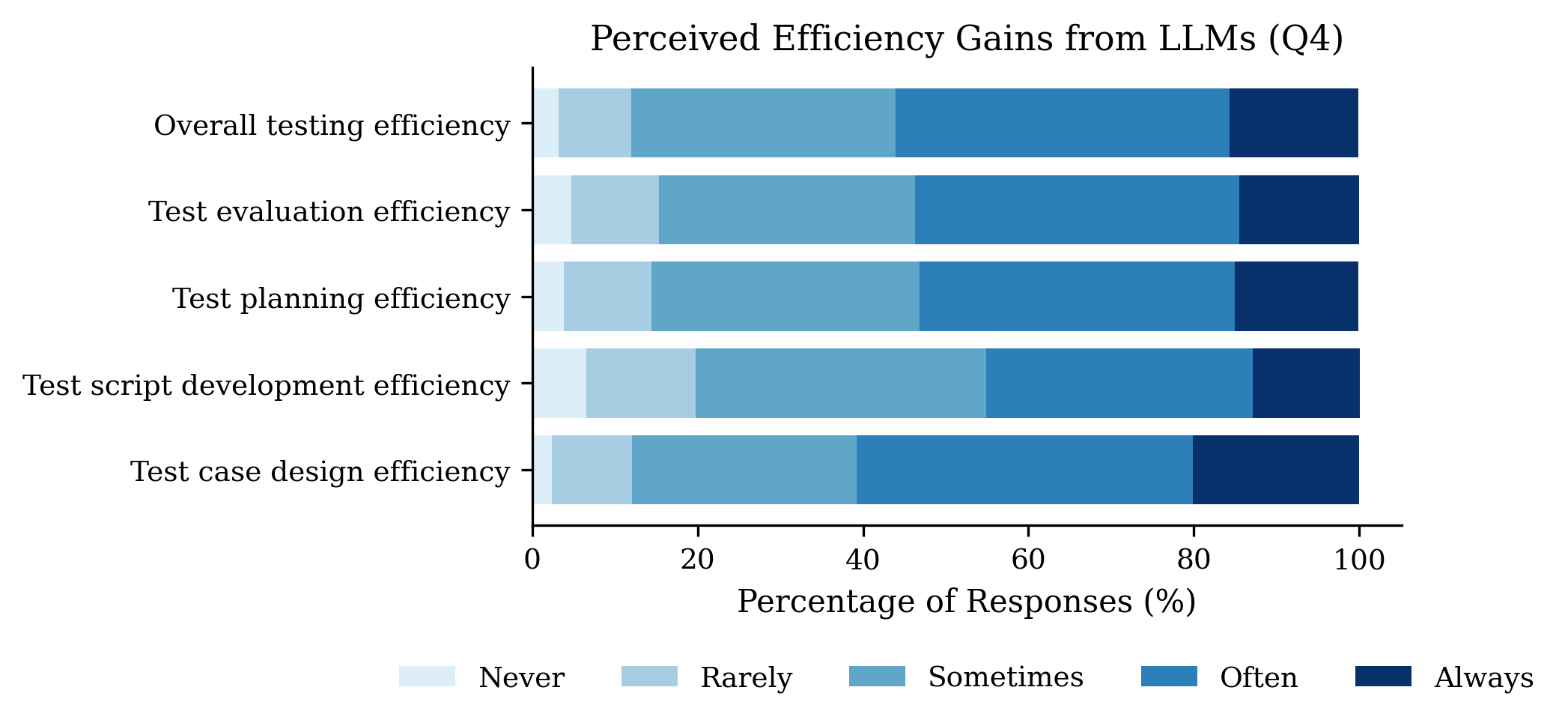}
    \caption{Perceived effectiveness of LLM assistance across software testing tasks, as reported in the large-scale competition survey.}
    \label{fig:q4}
\end{figure}

As shown in \textbf{Figure~\ref{fig:q4}}, students consistently perceived LLMs as more useful and reliable for text-oriented tasks, such as requirements analysis and test report generation.
Ratings for these tasks were relatively concentrated, suggesting more consistent perceived usefulness across respondents.

In contrast, perceived usefulness for automated test script generation exhibited substantially greater dispersion. We treat this dispersion as an indicator of heterogeneous user experiences in script generation, potentially reflecting differences in how respondents specified execution-relevant details (e.g., environment assumptions, element locators, synchronization logic, and verification conditions).

These perceptual patterns are consistent with the task-dependent prompt design difficulties identified in the classroom study (Figure~\ref{fig:Bubble}), where script generation prompts more frequently surfaced execution-sensitive under-specification and iterative clarification needs.
In addition to the competition survey, task-specific Likert-scale ratings were also collected from the 15 classroom participants using the same items. These within-subject responses are summarized in supplementary material.

To further characterize the effort associated with obtaining usable outputs in script generation, we next examine self-reported interaction rounds and debugging time as complementary process-level signals.

\subsubsection{Interaction Cost and the Illusion of Efficiency in Script Generation}

To further understand why automated test script generation emerged as a particularly challenging task, we analyzed students’ self-reported interaction costs and debugging overhead associated with LLM use.
Rather than focusing on output correctness, this analysis characterizes the \emph{process-level effort} required to obtain usable results.

Both interaction rounds and debugging time were collected via self-reported survey items.
The exact question wording and response options are listed in supplementary material. 

\begin{figure}[htbp]
    \centering
    \includegraphics[width=0.8\linewidth]{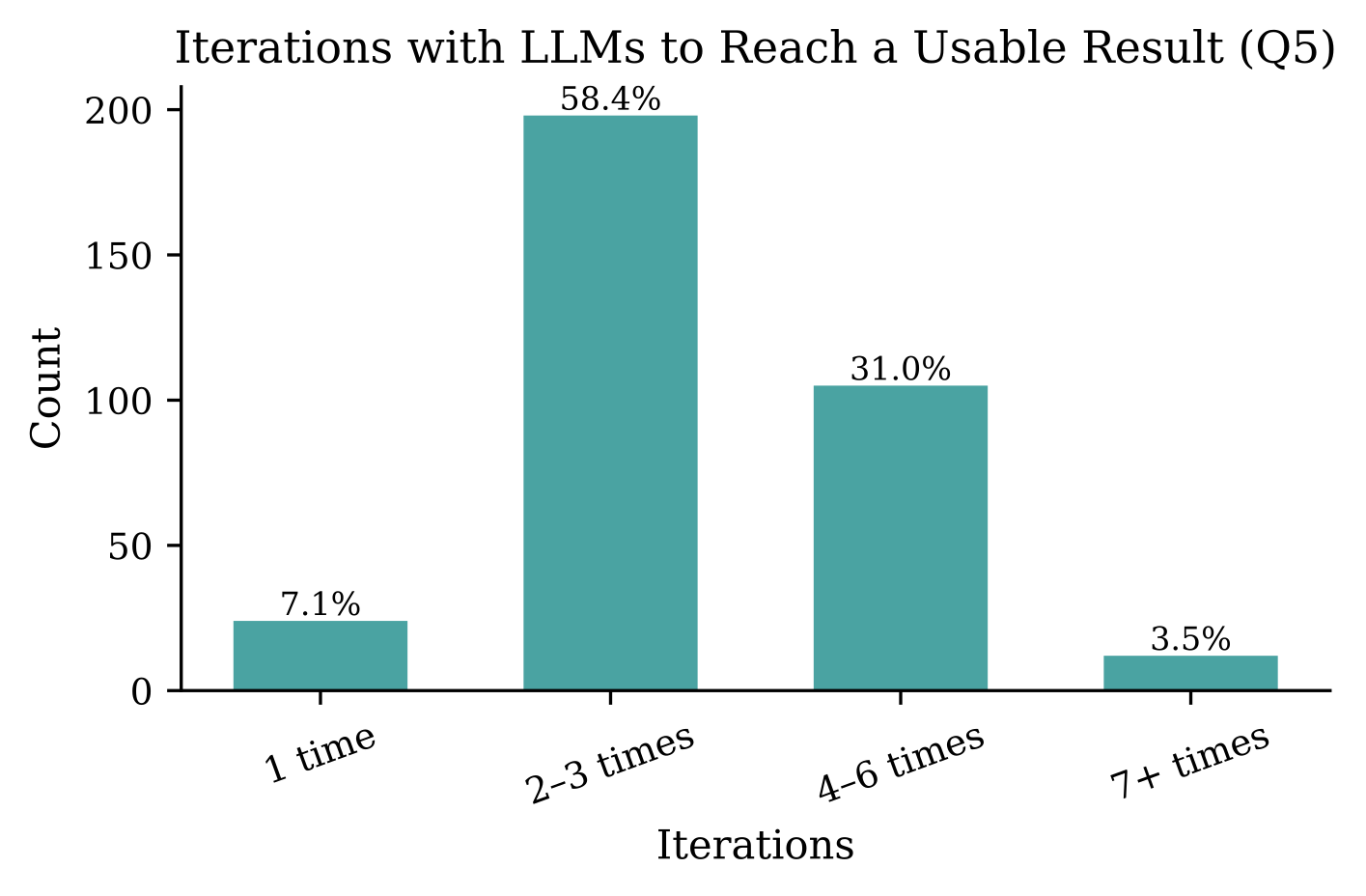}
    \caption{Reported interaction rounds required to obtain usable test scripts}
    \label{fig:q5_iterations}
\end{figure}

Survey responses indicate that script generation often required multiple rounds of interaction with LLMs.
As shown in \textbf{Figure~\ref{fig:q5_iterations}}, a majority of respondents reported engaging in two to three interaction rounds, while a substantial proportion required four or more iterations before obtaining a usable script. These repeated interactions suggest that initial LLM outputs frequently failed to meet execution or logic requirements, necessitating iterative clarification and correction. Note that “lack of interaction” in our coding does not mean fewer turns. Instead, it refers to the absence of proactive, strategy-driven iteration (e.g., decomposing the task, asking the model to confirm assumptions, and validating intermediate outputs). In script generation, students often started with a one-shot expectation and only iterated reactively after execution failures, which can lead to more turns but lower interaction quality.

\begin{figure}[htbp]
    \centering
    \includegraphics[width=0.8\linewidth]{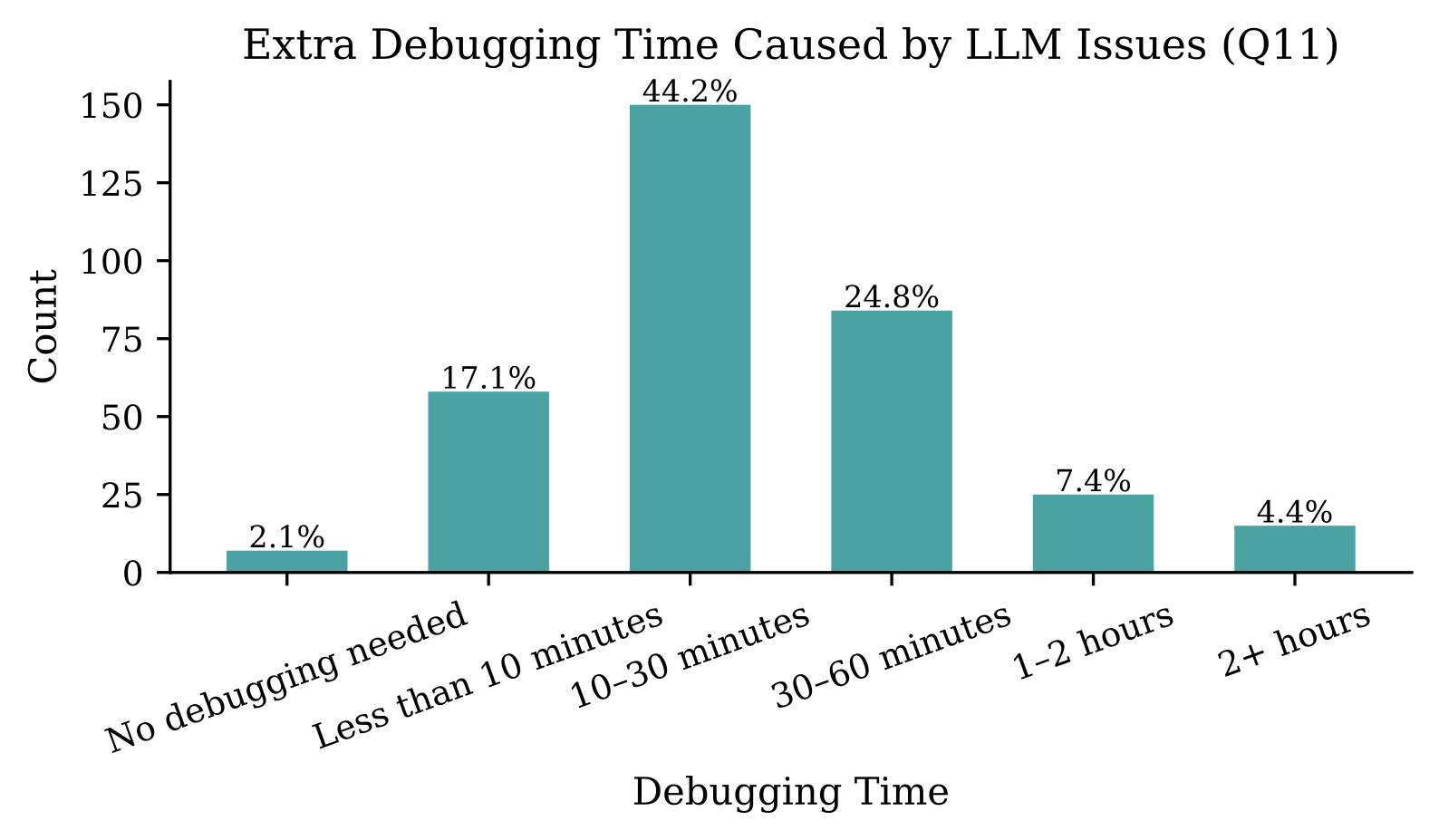}
    \caption{Self-reported additional debugging time incurred due to issues in LLM-generated scripts.}
    \label{fig:q11_debugtime}
\end{figure}

Beyond interaction rounds, students also reported considerable debugging overhead.
As illustrated in \textbf{Figure~\ref{fig:q11_debugtime}}, many respondents spent non-trivial additional time fixing issues in LLM-generated scripts, with some indicating that debugging time approached or even exceeded that required for manual implementation.
Open-ended responses further revealed recurring problems such as incompatible dependencies, inaccurate element locators, missing synchronization logic, and insufficient exception handling\cite{10.1145/3472673.3473967}.

Taken together, higher interaction rounds and increased debugging time co-occurred as a consistent pattern in script generation.
From a distributional perspective, these results describe a \emph{high-interaction-cost, low-certainty} experience: although LLMs are perceived as potentially useful, obtaining reliable and executable scripts often requires substantial additional effort.
This mismatch gives rise to an \emph{illusion of efficiency}~\cite{bender2020climbing}, where perceived usefulness does not translate into reduced workload or smoother task execution.

Importantly, these findings should not be interpreted as evidence that LLMs are inherently unsuitable for automated test script generation.
Rather, they indicate that, in the absence of instructional guidance, students struggle to externalize critical technical context and constraints during LLM interaction, leading to repeated revisions and downstream debugging.
This observation motivates an examination of whether such difficulties are \emph{pedagogically addressable}\cite{liu2024make}, rather than being solely attributable to model limitations, which we explore in Phase~II.

\subsubsection{Awareness of Prompt Quality and Learning Concerns}

Beyond task performance, the survey also revealed students’ reflective awareness of LLM limitations. Many respondents attributed script generation failures to missing critical information, particularly execution-relevant technical context (e.g., environment configuration and framework versions), element locator details, and underspecified business logic descriptions (see supplementary material).

Consistent with these attributions, respondents also linked failure experiences to prompt-quality issues such as vague task descriptions, incomplete context, and insufficient iterative refinement during LLM interaction (see supplementary material).
In addition to these process-related concerns, a substantial proportion of students expressed worries about over-reliance on LLMs, perceiving potential negative impacts on their test design skills, scripting abilities, and long-term professional development (see supplementary material).

These responses suggest that students are not passive consumers of LLM outputs. Instead, they actively weigh perceived efficiency gains against learning-related risks and skill development concerns. This implies that, in educational contexts, focusing solely on tool efficiency may be insufficient; instructional support should also address how students articulate task assumptions and engage in reflective, iterative use of LLMs.

Motivated by this observation, we next examine whether the identified difficulties are pedagogically addressable, that is, whether they can be meaningfully responded to through instructional guidance.

\paragraph{Summary of Phase I Findings.}
The results of Phase~I reveal a coherent picture of how task characteristics shape students’ learning-oriented use of LLMs in software testing.
Across tasks, students exhibited recurring prompt-related difficulties, but these difficulties were not uniformly distributed.
Text-oriented tasks such as requirements analysis and test case design showed relatively stable interaction patterns, with comparable distributions of prompt issues and more consistent perceived usefulness.
In contrast, automated test script generation emerged as a structurally different interaction context.
In this task, students were substantially less likely to engage in iterative refinement, perceived LLM assistance as less reliable, and incurred markedly higher interaction and debugging costs.
Importantly, these challenges did not stem from a single source, but reflected a misalignment between task demands and students’ ability to externalize execution context, constraints, and assumptions during LLM interaction.
Together, these findings suggest that difficulties in LLM-assisted software testing are not merely tool-related, but are deeply task-dependent and process-oriented, motivating an examination of whether such difficulties can be pedagogically addressed through instructional support.

\subsection{Phase II: Translating Identified Challenges into Instructional Support}

After identifying systematic challenges in LLM-assisted software testing, Phase~II explores through an illustrative practice whether these issues may be pedagogically addressable and whether they can be preliminarily translated into instructional support design considerations. We emphasize that Phase~II does not aim to propose or validate a new method, but rather to illustrate whether the issues identified in Phase~I can be pedagogically responded to.

Grounded in Phase~I findings, we focus on automated test script generation, which exhibited higher interaction costs and more unstable user experiences in both classroom artifacts and survey self-reports. In the context of this study, this task type was therefore selected for closer examination.

For reference, an illustrative version of the prompt scaffold used in Phase~II is provided in supplementary material. 

\subsubsection{Changes in Script Error Patterns}

Following the introduction of the instructional scaffold, we examined how students’ script issues manifested across different stages of the testing process, rather than treating errors as isolated technical failures. From a stage-aware perspective, several observed changes were associated with earlier testing stages, particularly goal formulation and context specification.

\begin{figure}[htbp]
    \centering
    \includegraphics[width=1\linewidth]{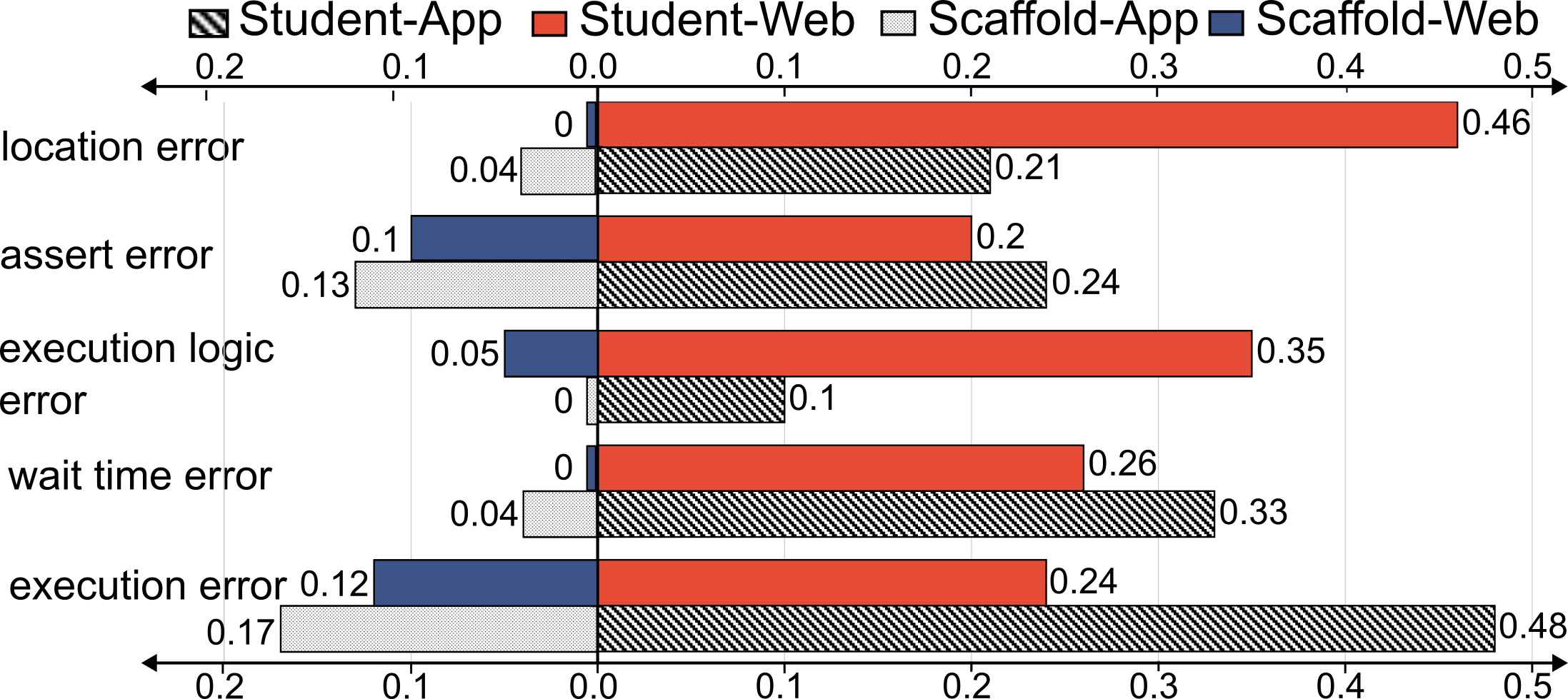}
    \caption{Illustrative distributions of script issue manifestations across different testing stages before and after the instructional practice. }

    \label{fig:his}
\end{figure}

As illustrated in Figure~\ref{fig:his}, issues related to insufficient environment specification, inaccurate element localization, and missing synchronization logic appeared less salient in the Phase~II artifacts. These categories correspond to testing stages where students must articulate assumptions about execution context and interaction timing before delegating actions to an LLM. Rather than indicating improved script quality, these observations suggest that students were more likely to externalize such assumptions during LLM interaction.

Importantly, we do not interpret these observations as evidence of reduced error rates or enhanced performance. Instead, they indicate a shift in how students engaged LLMs across testing stages specifically, a tendency to make stage-relevant constraints explicit rather than implicitly delegating them to the model.

\subsubsection{Platform-Specific Observations (Web vs.\ Mobile)}
When interpreted through a stage-aware lens, platform-specific patterns further contextualize how testing stages shape students’ LLM usage. In the web-testing artifacts, issues that typically emerge at the execution planning and synchronization stages—such as incomplete operation sequencing and missing wait conditions—appeared less salient after the scaffold was introduced. These issues are closely coupled with dynamic interface behaviors and timing assumptions, which are often left implicit when students rely primarily on screenshots or high-level descriptions during prompting.

For mobile-testing artifacts, the most noticeable shifts were concentrated at the context specification stage, particularly in how students articulated element localization. Rather than delegating locator selection to the LLM as an implementation detail, students more frequently externalized hierarchical or structural cues (e.g., UI containment relations) that support grounded locator construction. This pattern is consistent with Phase~I observations suggesting that mobile testing difficulties often arise from implicit assumptions about UI structure and element accessibility.

Given the illustrative role of Phase~II, we interpret these platform-specific patterns as descriptive indications of which testing stages become more salient under different platform characteristics, rather than as comparative claims about platform difficulty or evidence of instructional effectiveness.

\subsubsection{Summary of Phase II Observations}

Overall, Phase~II illustrates how some of the challenges identified in Phase~I may be pedagogically addressed by regulating students’ engagement with LLMs across testing stages. From a stage-aware perspective, the observed shifts suggest that students became more inclined to articulate goals, contextual constraints, and validation logic before or during LLM interaction, rather than delegating entire stages opportunistically.

These observations do not imply that the instructional scaffold resolves testing difficulties or improves performance outcomes. Instead, they demonstrate that challenges previously manifested as script execution failures can be reframed as learning issues related to stage-level reasoning and articulation. By making such reasoning visible, instructional support can create opportunities for reflection and discussion around LLM usage practices.

Taken together, Phase~II reinforces the central implication of Phase~I: in software testing education, the key challenge in integrating LLMs lies not in controlling access to the tools themselves, but in designing instructional structures that shape how and when learners engage with them throughout the testing process.

\section{Discussion}

This study examines students’ authentic learning behaviors when using LLMs in software testing education. By combining a classroom-based exploratory study with a competition survey, we systematically reveal task-dependent difficulties in LLM-assisted testing and discuss their educational implications. This section synthesizes the main findings from three perspectives: the nature of the problems, plausible explanatory mechanisms, and pedagogical implications.

\subsection{What Problems Do Students Systematically Encounter When Using LLMs?}

A primary finding of this study is that the difficulties students encounter when using LLMs for software testing tasks are not scattered or incidental; instead, they exhibit stable and generalizable patterns. Through analyses of students’ prompts and interaction processes, we identified six recurring types of issues: rigid usage patterns, missing context, lack of interaction, insufficient constraints, vague task descriptions, and missing input/output specifications.

These issues are not only surface-level deficiencies in prompt phrasing\cite{kazemitabaar2023novices}; they also reflect a potentially simplified understanding of how LLMs should be engaged during task work\cite{logacheva2024evaluating}. Notably, such understandings do not manifest uniformly across tasks. In our data, some students tended to treat LLMs as ``automatic answer generators,'' expecting a complete solution from a one-shot prompt, while overlooking that effective LLM use often involves an iterative process of clarifying goals, conditions, and assumptions. This cognitive tendency varied across testing tasks, suggesting that the emergence of prompt-related issues is not fully determined by individual ability alone, but is closely related to task characteristics.

Accordingly, these difficulties can be understood as an \emph{interaction mismatch} situated in a learning context, rather than merely technical or operational errors. This perspective helps shift attention from ``how to improve model outputs'' to ``how students understand and use the model.''

Viewed through a stage-aware lens, these recurring issues are not merely prompt-level deficiencies, but reflect systematic breakdowns in students’ reasoning at different stages of the testing process. For example, missing context and insufficient constraints often emerge during context specification, while lack of interaction is more closely associated with execution planning and validation stages. This perspective reframes prompt-related problems as stage-level articulation failures, rather than isolated interaction mistakes.

\subsection{Why Script Generation Emerges as a Persistent Bottleneck}

Across testing tasks, automated test script generation consistently appeared to involve higher uncertainty, greater interaction cost, and less stable user experiences, and was thus perceived by students as particularly challenging. This pattern was observed in both the classroom study and the survey\cite{LI2025103942}. Its underlying reasons can be interpreted from two angles: task-related cognitive load and the complexity of articulating constraints.

First, compared to requirements analysis or documentation tasks, script generation requires students to specify multiple layers of information simultaneously, including execution environments, page structures, operation sequences, synchronization mechanisms, and verification logic\cite{aghababaeyan2023black}. These elements demand both domain knowledge and precise articulation to the LLM. When any component is missing or underspecified, the generated scripts are more likely to be non-executable or logically fragile.

Second, students often view programming tasks as ``deterministic problems'' and therefore expect the model to produce correct code in a single attempt, which may reduce their willingness to iterate proactively. This expectation does not align well with the generative nature of LLMs. When outputs are incomplete or incorrect, students frequently resort to multiple rounds of prompting and substantial debugging before obtaining usable results. The resulting interaction cost and debugging burden make script generation the most frustrating task in students’ experiences.

These findings suggest that difficulties in script generation cannot be attributed entirely to model capability; they may also reflect students’ learning challenges in task decomposition, making conditions explicit, and diagnosing failures.

Importantly, the difficulty of script generation is not solely attributable to its technical complexity. Phase~II suggests that this task amplifies failures across multiple testing stages simultaneously. Students must coordinate goal formulation, context specification, execution planning, synchronization, and validation, often within a single interaction. When these stages are not explicitly articulated, students tend to delegate entire segments of the testing process to the LLM, resulting in fragile outputs and increased downstream effort.

\subsection{Educational Implications for Learning-Oriented LLM Use}

Based on the above analyses, we argue that the key to integrating LLMs into software testing education lies not in teaching specific tool tricks, but in repositioning LLM use as a \emph{learning activity}. Students should be guided to understand that high-quality LLM outputs are not ``automatically obtained,'' but depend on explicit articulation of testing intent, assumptions, and constraints.

Findings from Phase~II further suggest that pedagogical guidance may be most effective when it regulates students’ engagement with LLMs across testing stages, rather than uniformly encouraging or restricting LLM use. When students were prompted to externalize testing goals, contextual assumptions, and validation logic at stage-relevant moments, observable shifts emerged in how script-related issues manifested. These shifts do not indicate performance improvement, but illustrate that students’ interaction with LLMs can be reshaped from opportunistic delegation toward more reflective, stage-aware collaboration.

In practice, educators may consider a staged, scaffolded approach. Early in a course, LLMs can be introduced for text-oriented tasks to help students become familiar with interaction dynamics. When moving toward higher-complexity tasks such as script generation, structured prompts or reflective checklists may be used to guide students to gradually assume greater responsibility for articulating conditions and making judgments. Such an approach may help shift LLM use from simple result acquisition toward more reflective learning support.

\subsection{Reflections on Student Agency and Over-Reliance}

It is also noteworthy that students were not passive recipients of LLM outputs. Many participants demonstrated clear awareness of LLM limitations and explicitly expressed concerns that over-reliance could weaken their testing capabilities and long-term competitiveness \cite{selwyn2019should, denny2023can, bender2021dangers, kazemitabaar2023studying}. This reflective stance indicates that students exhibit agency in LLM use and actively weigh efficiency gains against potential learning risks\cite{mohammadkhani2023systematic}.

This finding has important implications for educators: the goal of instructional support should not be to restrict or replace LLM use, but to help students develop appropriate boundaries and judgment criteria. When students can recognize when to rely on LLMs and when to return to testing principles and manual analysis, LLMs are more likely to serve as learning-support tools rather than mere substitutes for independent thinking.

In summary, the educational value of LLMs in software testing does not depend solely on their generative capability; it depends on whether LLM use is embedded within learning-oriented and stage-aware instructional design that supports students in making informed decisions about when and how to engage with LLMs. By analyzing students’ authentic usage behaviors, this study provides empirical grounding for this shift in perspective.

\section{Threats to Validity}

\subsection{Internal Validity}

The classroom study involved a small sample from a single course, where peer influence and informal knowledge exchange may have shaped LLM interactions. To mitigate this, we triangulated multiple data sources—including independently submitted artifacts, LLM logs, classroom observations, and surveys—and focused on recurring patterns rather than isolated behaviors.

Students’ interactions were unguided by a strict protocol, introducing variability influenced by individual habits and prior LLM exposure. While this reduces experimental control, it captures authentic usage behaviors and supports ecological validity. Phase~II applied a lightweight instructional scaffold to examine how script-related issues were articulated across stages. This phase was not a controlled experiment, and observations should not be interpreted as evidence of instructional effectiveness, but rather as a means to explore whether Phase~I difficulties could be meaningfully addressed pedagogically.

\subsection{External Validity}

Participants in the classroom-based study were undergraduate software engineering students from a single institution, which may limit generalizability to other institutions, disciplines, or learner populations.

To partially address this limitation, we complemented the classroom study with a large-scale survey administered prior to a national software testing competition, drawing responses from participants across multiple universities. While this competition-oriented population does not represent a typical classroom cohort and may emphasize performance-oriented or tool-centric perspectives, it enables a cross-context examination of whether task-dependent patterns observed in a controlled course setting also appear in a broader and more diverse context. Nevertheless, the applicability of the findings to other instructional settings, educational levels, or software engineering subdomains requires further empirical investigation.

\subsection{Construct Validity}

Several findings rely on self-reported measures, such as perceived LLM effectiveness, interaction rounds, and debugging time, which may be affected by recall bias. We interpret these as approximate indicators of interaction patterns and emphasize relative trends rather than absolute values.

Qualitative analyses of prompts and interaction artifacts involve researcher interpretation. To reduce subjectivity, we focused on recurring issue categories, iteratively refined the coding scheme, and emphasized distributional patterns over marginal differences, supporting interpretive consistency while acknowledging inherent limitations of qualitative coding.

\section{Conclusion}

This paper presents an empirical investigation of students’ use of LLMs in software testing education from a learning-behavior perspective. We combine a classroom-based exploratory study involving 15 students with a large-scale competition survey (337 valid responses) to examine how students’ LLM-related difficulties vary across different testing tasks. Our findings reveal systematic, task-dependent patterns: text-oriented tasks tend to yield more consistent self-reported experiences, whereas automated test script generation is associated with higher numbers of interaction rounds, increased debugging effort, and more heterogeneous experiences. These contrasts motivate a stage-aware interpretation of where breakdowns occur during LLM-assisted work.

Analysis indicates that challenges in script generation are closely linked to students’ difficulty in making stage-relevant testing information explicit during LLM interaction—such as testing goals, execution context, constraints, synchronization assumptions, and validation intent—beyond inherent model limitations. From a learning-oriented perspective, these breakdowns highlight pedagogical opportunities: supporting students in articulating their testing reasoning can shift LLM use from ad-hoc delegation toward more reflective, collaborative problem solving.

We illustrate these insights through a classroom vignette that operationalizes the observed breakdowns into a lightweight, stage-aware prompt scaffold for script generation. Descriptive analysis shows shifts in how students externalize testing-relevant information across stages (e.g., context specification), underscoring that the educational value of LLMs depends not solely on generative capability, but on whether their integration is guided by stage-aware instructional design.

\begin{acks}
This work is supported by the Scientific Research Platforms and Projects of Guangdong Provincial Education Department under Grant 2024ZDZX1068; the National Key Research and Development Program of China (Research and Demonstration Application of Key Technologies for Personalized Learning Driven by Educational Big Data) under Grant 2023YFC3341200; the National Key R\&D Program of China Key Special Project under Grant 2023YFC3341204; the Tertiary Education Scientific Research Project of Guangzhou Municipal Education Bureau under Grant 2024312300; and the Collaborative Innovation Center for Intelligent Educational Technology of Guangzhou under Grants 2023B04J0007 and 2025B04J0007.
\end{acks}

\bibliographystyle{ACM-Reference-Format} 
\bibliography{main.bib}


\begin{thebibliography}{40}


\ifx \showCODEN    \undefined \def \showCODEN     #1{\unskip}     \fi
\ifx \showISBNx    \undefined \def \showISBNx     #1{\unskip}     \fi
\ifx \showISBNxiii \undefined \def \showISBNxiii  #1{\unskip}     \fi
\ifx \showISSN     \undefined \def \showISSN      #1{\unskip}     \fi
\ifx \showLCCN     \undefined \def \showLCCN      #1{\unskip}     \fi
\ifx \shownote     \undefined \def \shownote      #1{#1}          \fi
\ifx \showarticletitle \undefined \def \showarticletitle #1{#1}   \fi
\ifx \showURL      \undefined \def \showURL       {\relax}        \fi
\providecommand\bibfield[2]{#2}
\providecommand\bibinfo[2]{#2}
\providecommand\natexlab[1]{#1}
\providecommand\showeprint[2][]{arXiv:#2}

\bibitem[Aghababaeyan et~al\mbox{.}(2023)]%
        {aghababaeyan2023black}
\bibfield{author}{\bibinfo{person}{Zohreh Aghababaeyan}, \bibinfo{person}{Manel Abdellatif}, \bibinfo{person}{Lionel Briand}, \bibinfo{person}{Ramesh S}, {and} \bibinfo{person}{Mojtaba Bagherzadeh}.} \bibinfo{year}{2023}\natexlab{}.
\newblock \showarticletitle{Black-Box Testing of Deep Neural Networks through Test Case Diversity}.
\newblock \bibinfo{journal}{\emph{IEEE Transactions on Software Engineering}} \bibinfo{volume}{49}, \bibinfo{number}{5} (\bibinfo{year}{2023}), \bibinfo{pages}{3182--3204}.
\newblock
\href{https://doi.org/10.1109/TSE.2023.3243522}{doi:\nolinkurl{10.1109/TSE.2023.3243522}}


\bibitem[Amalfitano et~al\mbox{.}(2023)]%
        {amalfitano2023artificial}
\bibfield{author}{\bibinfo{person}{Domenico Amalfitano}, \bibinfo{person}{Stefano Faralli}, \bibinfo{person}{Jean Carlo~Rossa Hauck}, \bibinfo{person}{Santiago Matalonga}, {and} \bibinfo{person}{Damiano Distante}.} \bibinfo{year}{2023}\natexlab{}.
\newblock \showarticletitle{Artificial intelligence applied to software testing: A tertiary study}.
\newblock \bibinfo{journal}{\emph{Comput. Surveys}} \bibinfo{volume}{56}, \bibinfo{number}{3} (\bibinfo{year}{2023}), \bibinfo{pages}{1--38}.
\newblock


\bibitem[Balse et~al\mbox{.}(2023)]%
        {balse2023exploring}
\bibfield{author}{\bibinfo{person}{Rishabh Balse}, \bibinfo{person}{Prajish Prasad}, {and} \bibinfo{person}{Jayakrishnan~Madathil Warriem}.} \bibinfo{year}{2023}\natexlab{}.
\newblock \showarticletitle{Exploring the Potential of GPT-4 in Automated Mentoring for Programming Courses}. In \bibinfo{booktitle}{\emph{Proceedings of the ACM Conference on Global Computing Education Vol 2}} (Hyderabad, India) \emph{(\bibinfo{series}{CompEd 2023})}. \bibinfo{publisher}{Association for Computing Machinery}, \bibinfo{address}{New York, NY, USA}, \bibinfo{pages}{191}.
\newblock
\showISBNx{9798400703744}
\href{https://doi.org/10.1145/3617650.3624946}{doi:\nolinkurl{10.1145/3617650.3624946}}


\bibitem[Bender et~al\mbox{.}(2021)]%
        {bender2021dangers}
\bibfield{author}{\bibinfo{person}{Emily~M. Bender}, \bibinfo{person}{Timnit Gebru}, \bibinfo{person}{Angelina McMillan-Major}, {and} \bibinfo{person}{Shmargaret Shmitchell}.} \bibinfo{year}{2021}\natexlab{}.
\newblock \showarticletitle{On the Dangers of Stochastic Parrots: Can Language Models Be Too Big?}. In \bibinfo{booktitle}{\emph{Proceedings of the 2021 ACM Conference on Fairness, Accountability, and Transparency}} (Virtual Event, Canada) \emph{(\bibinfo{series}{FAccT '21})}. \bibinfo{publisher}{Association for Computing Machinery}, \bibinfo{address}{New York, NY, USA}, \bibinfo{pages}{610–623}.
\newblock
\showISBNx{9781450383097}
\href{https://doi.org/10.1145/3442188.3445922}{doi:\nolinkurl{10.1145/3442188.3445922}}


\bibitem[Bender and Koller(2020)]%
        {bender2020climbing}
\bibfield{author}{\bibinfo{person}{Emily~M Bender} {and} \bibinfo{person}{Alexander Koller}.} \bibinfo{year}{2020}\natexlab{}.
\newblock \showarticletitle{Climbing towards NLU: On meaning, form, and understanding in the age of data}. In \bibinfo{booktitle}{\emph{Proceedings of the 58th annual meeting of the association for computational linguistics}}. \bibinfo{pages}{5185--5198}.
\newblock


\bibitem[Bommasani(2021)]%
        {bommasani2021opportunities}
\bibfield{author}{\bibinfo{person}{Rishi Bommasani}.} \bibinfo{year}{2021}\natexlab{}.
\newblock \showarticletitle{On the opportunities and risks of foundation models}.
\newblock \bibinfo{journal}{\emph{arXiv preprint arXiv:2108.07258}} (\bibinfo{year}{2021}).
\newblock


\bibitem[Cheiran et~al\mbox{.}(2017)]%
        {cheiran2017problem}
\bibfield{author}{\bibinfo{person}{Jean Felipe~P Cheiran}, \bibinfo{person}{Elder de M.~Rodrigues}, \bibinfo{person}{Ewerson~Luiz de S.~Carvalho}, {and} \bibinfo{person}{Jo{\~a}o Pablo~S da Silva}.} \bibinfo{year}{2017}\natexlab{}.
\newblock \showarticletitle{Problem-based learning to align theory and practice in software testing teaching}. In \bibinfo{booktitle}{\emph{Proceedings of the XXXI Brazilian Symposium on Software Engineering}}. \bibinfo{pages}{328--337}.
\newblock


\bibitem[Clegg et~al\mbox{.}(2017)]%
        {clegg2017mut}
\bibfield{author}{\bibinfo{person}{Benjamin~S Clegg}, \bibinfo{person}{Jos{\'e}~Miguel Rojas}, {and} \bibinfo{person}{Gordon Fraser}.} \bibinfo{year}{2017}\natexlab{}.
\newblock \showarticletitle{Teaching software testing concepts using a mutation testing game}. In \bibinfo{booktitle}{\emph{2017 IEEE/ACM 39th International Conference on Software Engineering: Software Engineering Education and Training Track (ICSE-SEET)}}. IEEE, \bibinfo{pages}{33--36}.
\newblock


\bibitem[Daun and Brings(2023)]%
        {daun2023chatgpt}
\bibfield{author}{\bibinfo{person}{Marian Daun} {and} \bibinfo{person}{Jennifer Brings}.} \bibinfo{year}{2023}\natexlab{}.
\newblock \showarticletitle{How ChatGPT will change software engineering education}. In \bibinfo{booktitle}{\emph{Proceedings of the 2023 Conference on Innovation and Technology in Computer Science Education V. 1}}. \bibinfo{pages}{110--116}.
\newblock


\bibitem[de~Souza et~al\mbox{.}(2014)]%
        {souza2014assess}
\bibfield{author}{\bibinfo{person}{Draylson~M de Souza}, \bibinfo{person}{Bruno~H Oliveira}, \bibinfo{person}{Jose~Carlos Maldonado}, \bibinfo{person}{Simone~RS Souza}, {and} \bibinfo{person}{Ellen~F Barbosa}.} \bibinfo{year}{2014}\natexlab{}.
\newblock \showarticletitle{Towards the use of an automatic assessment system in the teaching of software testing}. In \bibinfo{booktitle}{\emph{2014 IEEE Frontiers in Education Conference (FIE) Proceedings}}. IEEE, \bibinfo{pages}{1--8}.
\newblock


\bibitem[Delgado-Pérez et~al\mbox{.}(2023)]%
        {delgado2022interevo}
\bibfield{author}{\bibinfo{person}{Pedro Delgado-Pérez}, \bibinfo{person}{Aurora Ramírez}, \bibinfo{person}{Kevin~J. Valle-Gómez}, \bibinfo{person}{Inmaculada Medina-Bulo}, {and} \bibinfo{person}{José~Raúl Romero}.} \bibinfo{year}{2023}\natexlab{}.
\newblock \showarticletitle{InterEvo-TR: Interactive Evolutionary Test Generation With Readability Assessment}.
\newblock \bibinfo{journal}{\emph{IEEE Transactions on Software Engineering}} \bibinfo{volume}{49}, \bibinfo{number}{4} (\bibinfo{year}{2023}), \bibinfo{pages}{2580--2596}.
\newblock
\href{https://doi.org/10.1109/TSE.2022.3227418}{doi:\nolinkurl{10.1109/TSE.2022.3227418}}


\bibitem[Denny et~al\mbox{.}(2023)]%
        {denny2023can}
\bibfield{author}{\bibinfo{person}{Paul Denny}, \bibinfo{person}{Hassan Khosravi}, \bibinfo{person}{Arto Hellas}, \bibinfo{person}{Juho Leinonen}, {and} \bibinfo{person}{Sami Sarsa}.} \bibinfo{year}{2023}\natexlab{}.
\newblock \showarticletitle{Can we trust AI-generated educational content? comparative analysis of human and AI-generated learning resources}.
\newblock \bibinfo{journal}{\emph{arXiv preprint arXiv:2306.10509}} (\bibinfo{year}{2023}).
\newblock


\bibitem[Eisty et~al\mbox{.}(2025)]%
        {eisty2025testing}
\bibfield{author}{\bibinfo{person}{Nasir~U Eisty}, \bibinfo{person}{Upulee Kanewala}, {and} \bibinfo{person}{Jeffrey~C Carver}.} \bibinfo{year}{2025}\natexlab{}.
\newblock \showarticletitle{Testing research software: an in-depth survey of practices, methods, and tools}.
\newblock \bibinfo{journal}{\emph{Empirical Software Engineering}} \bibinfo{volume}{30}, \bibinfo{number}{3} (\bibinfo{year}{2025}), \bibinfo{pages}{81}.
\newblock


\bibitem[Elgrably and Oliveira(2022)]%
        {elgrably2022quasi}
\bibfield{author}{\bibinfo{person}{Isaac Elgrably} {and} \bibinfo{person}{Sandro Oliveira}.} \bibinfo{year}{2022}\natexlab{}.
\newblock \showarticletitle{A Quasi-Experimental Evaluation of Teaching Software Testing in Software Quality Assurance Subject during a Post-Graduate Computer Science Course}.
\newblock \bibinfo{journal}{\emph{International Journal of Emerging Technologies in Learning (iJET)}} \bibinfo{volume}{17}, \bibinfo{number}{5} (\bibinfo{year}{2022}), \bibinfo{pages}{57--86}.
\newblock


\bibitem[Jalil et~al\mbox{.}(2023)]%
        {jalil2023chatgpt}
\bibfield{author}{\bibinfo{person}{Sajed Jalil}, \bibinfo{person}{Suzzana Rafi}, \bibinfo{person}{Thomas~D. LaToza}, \bibinfo{person}{Kevin Moran}, {and} \bibinfo{person}{Wing Lam}.} \bibinfo{year}{2023}\natexlab{}.
\newblock \showarticletitle{ChatGPT and Software Testing Education: Promises \& Perils}. In \bibinfo{booktitle}{\emph{2023 IEEE International Conference on Software Testing, Verification and Validation Workshops (ICSTW)}}. \bibinfo{pages}{4130--4137}.
\newblock
\href{https://doi.org/10.1109/ICSTW58534.2023.00078}{doi:\nolinkurl{10.1109/ICSTW58534.2023.00078}}


\bibitem[Kazemitabaar et~al\mbox{.}(2023a)]%
        {kazemitabaar2023studying}
\bibfield{author}{\bibinfo{person}{Majeed Kazemitabaar}, \bibinfo{person}{Justin Chow}, \bibinfo{person}{Carl Ka~To Ma}, \bibinfo{person}{Barbara~J Ericson}, \bibinfo{person}{David Weintrop}, {and} \bibinfo{person}{Tovi Grossman}.} \bibinfo{year}{2023}\natexlab{a}.
\newblock \showarticletitle{Studying the effect of AI code generators on supporting novice learners in introductory programming}. In \bibinfo{booktitle}{\emph{Proceedings of the 2023 CHI conference on human factors in computing systems}}. \bibinfo{pages}{1--23}.
\newblock


\bibitem[Kazemitabaar et~al\mbox{.}(2023b)]%
        {kazemitabaar2023novices}
\bibfield{author}{\bibinfo{person}{Majeed Kazemitabaar}, \bibinfo{person}{Xinying Hou}, \bibinfo{person}{Austin Henley}, \bibinfo{person}{Barbara~Jane Ericson}, \bibinfo{person}{David Weintrop}, {and} \bibinfo{person}{Tovi Grossman}.} \bibinfo{year}{2023}\natexlab{b}.
\newblock \showarticletitle{How novices use LLM-based code generators to solve CS1 coding tasks in a self-paced learning environment}. In \bibinfo{booktitle}{\emph{Proceedings of the 23rd Koli calling international conference on computing education research}}. \bibinfo{pages}{1--12}.
\newblock


\bibitem[Kokotsaki et~al\mbox{.}(2016)]%
        {kokotsaki2016project}
\bibfield{author}{\bibinfo{person}{Dimitra Kokotsaki}, \bibinfo{person}{Victoria Menzies}, {and} \bibinfo{person}{Andy Wiggins}.} \bibinfo{year}{2016}\natexlab{}.
\newblock \showarticletitle{Project-based learning: A review of the literature}.
\newblock \bibinfo{journal}{\emph{Improving schools}} \bibinfo{volume}{19}, \bibinfo{number}{3} (\bibinfo{year}{2016}), \bibinfo{pages}{267--277}.
\newblock


\bibitem[Koutcheme et~al\mbox{.}(2024)]%
        {koutcheme2024using}
\bibfield{author}{\bibinfo{person}{Charles Koutcheme}, \bibinfo{person}{Nicola Dainese}, {and} \bibinfo{person}{Arto Hellas}.} \bibinfo{year}{2024}\natexlab{}.
\newblock \showarticletitle{Using Program Repair as a Proxy for Language Models’ Feedback Ability in Programming Education}. In \bibinfo{booktitle}{\emph{Workshop on Innovative Use of NLP for Building Educational Applications}}. Association for Computational Linguistics, \bibinfo{pages}{165--181}.
\newblock


\bibitem[Li et~al\mbox{.}(2025)]%
        {LI2025103942}
\bibfield{author}{\bibinfo{person}{Yihao Li}, \bibinfo{person}{Pan Liu}, \bibinfo{person}{Haiyang Wang}, \bibinfo{person}{Jie Chu}, {and} \bibinfo{person}{W.~Eric Wong}.} \bibinfo{year}{2025}\natexlab{}.
\newblock \showarticletitle{Evaluating large language models for software testing}.
\newblock \bibinfo{journal}{\emph{Computer Standards \& Interfaces}}  \bibinfo{volume}{93} (\bibinfo{year}{2025}), \bibinfo{pages}{103942}.
\newblock
\showISSN{0920-5489}
\href{https://doi.org/10.1016/j.csi.2024.103942}{doi:\nolinkurl{10.1016/j.csi.2024.103942}}


\bibitem[Liu et~al\mbox{.}(2024)]%
        {liu2024make}
\bibfield{author}{\bibinfo{person}{Zhe Liu}, \bibinfo{person}{Chunyang Chen}, \bibinfo{person}{Junjie Wang}, \bibinfo{person}{Mengzhuo Chen}, \bibinfo{person}{Boyu Wu}, \bibinfo{person}{Xing Che}, \bibinfo{person}{Dandan Wang}, {and} \bibinfo{person}{Qing Wang}.} \bibinfo{year}{2024}\natexlab{}.
\newblock \showarticletitle{Make llm a testing expert: Bringing human-like interaction to mobile gui testing via functionality-aware decisions}. In \bibinfo{booktitle}{\emph{Proceedings of the IEEE/ACM 46th International Conference on Software Engineering}}. \bibinfo{pages}{1--13}.
\newblock


\bibitem[Logacheva et~al\mbox{.}(2024)]%
        {logacheva2024evaluating}
\bibfield{author}{\bibinfo{person}{Evanfiya Logacheva}, \bibinfo{person}{Arto Hellas}, \bibinfo{person}{James Prather}, \bibinfo{person}{Sami Sarsa}, {and} \bibinfo{person}{Juho Leinonen}.} \bibinfo{year}{2024}\natexlab{}.
\newblock \showarticletitle{Evaluating contextually personalized programming exercises created with generative AI}. In \bibinfo{booktitle}{\emph{Proceedings of the 2024 ACM Conference on International Computing Education Research-Volume 1}}. \bibinfo{pages}{95--113}.
\newblock


\bibitem[MacNeil et~al\mbox{.}(2024)]%
        {macneil2024synthetic}
\bibfield{author}{\bibinfo{person}{Stephen MacNeil}, \bibinfo{person}{Magdalena Rogalska}, \bibinfo{person}{Juho Leinonen}, \bibinfo{person}{Paul Denny}, \bibinfo{person}{Arto Hellas}, {and} \bibinfo{person}{Xandria Crosland}.} \bibinfo{year}{2024}\natexlab{}.
\newblock \showarticletitle{Synthetic Students: A Comparative Study of Bug Distribution Between Large Language Models and Computing Students}. In \bibinfo{booktitle}{\emph{Proceedings of the 2024 on ACM Virtual Global Computing Education Conference V. 1}}. \bibinfo{pages}{137--143}.
\newblock


\bibitem[Majumdar et~al\mbox{.}(2024)]%
        {majumdar2024mining}
\bibfield{author}{\bibinfo{person}{Rwitajit Majumdar}, \bibinfo{person}{Prajish Prasad}, {and} \bibinfo{person}{Aamod Sane}.} \bibinfo{year}{2024}\natexlab{}.
\newblock \showarticletitle{Mining Epistemic Actions of Programming Problem Solving with Chat-GPT}. In \bibinfo{booktitle}{\emph{Proceedings of the 17th International Conference on Educational Data Mining}}. \bibinfo{pages}{628--633}.
\newblock


\bibitem[Michaeli and Romeike(2017)]%
        {michaeli2017test}
\bibfield{author}{\bibinfo{person}{Tilman Michaeli} {and} \bibinfo{person}{Ralf Romeike}.} \bibinfo{year}{2017}\natexlab{}.
\newblock \showarticletitle{Addressing teaching practices regarding software quality: Testing and debugging in the classroom}. In \bibinfo{booktitle}{\emph{Proceedings of the 12th Workshop on Primary and Secondary Computing Education}}. \bibinfo{pages}{105--106}.
\newblock


\bibitem[Mohammadkhani et~al\mbox{.}(2023)]%
        {mohammadkhani2023systematic}
\bibfield{author}{\bibinfo{person}{Ahmad~Haji Mohammadkhani}, \bibinfo{person}{Nitin~Sai Bommi}, \bibinfo{person}{Mariem Daboussi}, \bibinfo{person}{Onkar Sabnis}, \bibinfo{person}{Chakkrit Tantithamthavorn}, {and} \bibinfo{person}{Hadi Hemmati}.} \bibinfo{year}{2023}\natexlab{}.
\newblock \showarticletitle{A systematic literature review of explainable AI for software engineering}.
\newblock \bibinfo{journal}{\emph{arXiv preprint arXiv:2302.06065}} (\bibinfo{year}{2023}).
\newblock


\bibitem[Neumann et~al\mbox{.}(2025)]%
        {DBLP:journals/te/NeumannYSDJ25}
\bibfield{author}{\bibinfo{person}{Alexander~Tobias Neumann}, \bibinfo{person}{Yue Yin}, \bibinfo{person}{Sulayman~K. Sowe}, \bibinfo{person}{Stefan Decker}, {and} \bibinfo{person}{Matthias Jarke}.} \bibinfo{year}{2025}\natexlab{}.
\newblock \showarticletitle{An LLM-Driven Chatbot in Higher Education for Databases and Information Systems}.
\newblock \bibinfo{journal}{\emph{IEEE Transactions on Education}} \bibinfo{volume}{68}, \bibinfo{number}{1} (\bibinfo{year}{2025}), \bibinfo{pages}{103--116}.
\newblock
\href{https://doi.org/10.1109/TE.2024.3467912}{doi:\nolinkurl{10.1109/TE.2024.3467912}}


\bibitem[Prather et~al\mbox{.}(2023)]%
        {prather2023robots}
\bibfield{author}{\bibinfo{person}{James Prather}, \bibinfo{person}{Paul Denny}, \bibinfo{person}{Juho Leinonen}, \bibinfo{person}{Brett~A Becker}, \bibinfo{person}{Ibrahim Albluwi}, \bibinfo{person}{Michelle Craig}, \bibinfo{person}{Hieke Keuning}, \bibinfo{person}{Natalie Kiesler}, \bibinfo{person}{Tobias Kohn}, \bibinfo{person}{Andrew Luxton-Reilly}, {et~al\mbox{.}}} \bibinfo{year}{2023}\natexlab{}.
\newblock \showarticletitle{The robots are here: Navigating the generative ai revolution in computing education}.
\newblock In \bibinfo{booktitle}{\emph{Proceedings of the 2023 working group reports on innovation and technology in computer science education}}. \bibinfo{pages}{108--159}.
\newblock


\bibitem[Selwyn(2019)]%
        {selwyn2019should}
\bibfield{author}{\bibinfo{person}{Neil Selwyn}.} \bibinfo{year}{2019}\natexlab{}.
\newblock \bibinfo{booktitle}{\emph{Should robots replace teachers?: AI and the future of education}}.
\newblock \bibinfo{publisher}{John Wiley \& Sons}.
\newblock


\bibitem[Sengul et~al\mbox{.}(2024)]%
        {sengul2024software}
\bibfield{author}{\bibinfo{person}{Cigdem Sengul}, \bibinfo{person}{Rumyana Neykova}, {and} \bibinfo{person}{Giuseppe Destefanis}.} \bibinfo{year}{2024}\natexlab{}.
\newblock \showarticletitle{Software engineering education in the era of conversational AI: current trends and future directions}.
\newblock \bibinfo{journal}{\emph{Frontiers in Artificial Intelligence}}  \bibinfo{volume}{7} (\bibinfo{year}{2024}), \bibinfo{pages}{1436350}.
\newblock


\bibitem[Sindre(2023)]%
        {Sindre_2023}
\bibfield{author}{\bibinfo{person}{Guttorm Sindre}.} \bibinfo{year}{2023}\natexlab{}.
\newblock \showarticletitle{AI Technology: Threats and Opportunities for Assessment Integrity in Introductory Programming}.
\newblock \bibinfo{journal}{\emph{Norsk IKT-konferanse for forskning og utdanning}} \bibinfo{number}{4} (\bibinfo{date}{Nov.} \bibinfo{year}{2023}).
\newblock
\urldef\tempurl%
\url{https://www.ntnu.no/ojs/index.php/nikt/article/view/5710}
\showURL{%
\tempurl}


\bibitem[Smith and Praphamontripong(2021)]%
        {10.1145/3472673.3473967}
\bibfield{author}{\bibinfo{person}{Chloe Smith} {and} \bibinfo{person}{Upsorn Praphamontripong}.} \bibinfo{year}{2021}\natexlab{}.
\newblock \showarticletitle{Analysis of the transition to a virtual learning semester in a college software testing course}. In \bibinfo{booktitle}{\emph{Proceedings of the 3rd International Workshop on Education through Advanced Software Engineering and Artificial Intelligence}} (Athens, Greece) \emph{(\bibinfo{series}{EASEAI 2021})}. \bibinfo{publisher}{Association for Computing Machinery}, \bibinfo{address}{New York, NY, USA}, \bibinfo{pages}{58–61}.
\newblock
\showISBNx{9781450386241}
\href{https://doi.org/10.1145/3472673.3473967}{doi:\nolinkurl{10.1145/3472673.3473967}}


\bibitem[Tang et~al\mbox{.}(2024)]%
        {tang2024chatgpt}
\bibfield{author}{\bibinfo{person}{Yutian Tang}, \bibinfo{person}{Zhijie Liu}, \bibinfo{person}{Zhichao Zhou}, {and} \bibinfo{person}{Xiapu Luo}.} \bibinfo{year}{2024}\natexlab{}.
\newblock \showarticletitle{Chatgpt vs sbst: A comparative assessment of unit test suite generation}.
\newblock \bibinfo{journal}{\emph{IEEE Transactions on Software Engineering}} \bibinfo{volume}{50}, \bibinfo{number}{6} (\bibinfo{year}{2024}), \bibinfo{pages}{1340--1359}.
\newblock


\bibitem[Towey and Chen(2015)]%
        {towey2015teaching}
\bibfield{author}{\bibinfo{person}{Dave Towey} {and} \bibinfo{person}{Tsong~Yueh Chen}.} \bibinfo{year}{2015}\natexlab{}.
\newblock \showarticletitle{Teaching software testing skills: Metamorphic testing as vehicle for creativity and effectiveness in software testing}. In \bibinfo{booktitle}{\emph{2015 IEEE International Conference on Teaching, Assessment, and Learning for Engineering (TALE)}}. IEEE, \bibinfo{pages}{161--162}.
\newblock


\bibitem[Vierhauser et~al\mbox{.}(2024)]%
        {vierhauser2024towards}
\bibfield{author}{\bibinfo{person}{Michael Vierhauser}, \bibinfo{person}{Iris Groher}, \bibinfo{person}{Tobias Antensteiner}, {and} \bibinfo{person}{Clemens Sauerwein}.} \bibinfo{year}{2024}\natexlab{}.
\newblock \showarticletitle{Towards Integrating Emerging AI Applications in SE Education}.
\newblock \bibinfo{journal}{\emph{arXiv preprint arXiv:2405.18062}} (\bibinfo{year}{2024}).
\newblock


\bibitem[Wang et~al\mbox{.}(2024)]%
        {wang2024software}
\bibfield{author}{\bibinfo{person}{Junjie Wang}, \bibinfo{person}{Yuchao Huang}, \bibinfo{person}{Chunyang Chen}, \bibinfo{person}{Zhe Liu}, \bibinfo{person}{Song Wang}, {and} \bibinfo{person}{Qing Wang}.} \bibinfo{year}{2024}\natexlab{}.
\newblock \showarticletitle{Software testing with large language models: Survey, landscape, and vision}.
\newblock \bibinfo{journal}{\emph{IEEE Transactions on Software Engineering}} \bibinfo{volume}{50}, \bibinfo{number}{4} (\bibinfo{year}{2024}), \bibinfo{pages}{911--936}.
\newblock


\bibitem[Wieser et~al\mbox{.}(2023)]%
        {wieser2023investigating}
\bibfield{author}{\bibinfo{person}{Markus Wieser}, \bibinfo{person}{Klaus Sch{\"o}ffmann}, \bibinfo{person}{Daniela Stefanics}, \bibinfo{person}{Andreas Bollin}, {and} \bibinfo{person}{Stefan Pasterk}.} \bibinfo{year}{2023}\natexlab{}.
\newblock \showarticletitle{Investigating the Role of ChatGPT in Supporting Text-Based Programming Education for Students and Teachers}. In \bibinfo{booktitle}{\emph{International Conference on Informatics in Schools: Situation, Evolution, and Perspectives}}. Springer Nature Switzerland Cham, \bibinfo{pages}{40--53}.
\newblock


\bibitem[Xue et~al\mbox{.}(2024)]%
        {xue2024llm4fin}
\bibfield{author}{\bibinfo{person}{Zhiyi Xue}, \bibinfo{person}{Liangguo Li}, \bibinfo{person}{Senyue Tian}, \bibinfo{person}{Xiaohong Chen}, \bibinfo{person}{Pingping Li}, \bibinfo{person}{Liangyu Chen}, \bibinfo{person}{Tingting Jiang}, {and} \bibinfo{person}{Min Zhang}.} \bibinfo{year}{2024}\natexlab{}.
\newblock \showarticletitle{Llm4fin: Fully automating llm-powered test case generation for fintech software acceptance testing}. In \bibinfo{booktitle}{\emph{Proceedings of the 33rd ACM SIGSOFT International Symposium on Software Testing and Analysis}}. \bibinfo{pages}{1643--1655}.
\newblock


\bibitem[Yu et~al\mbox{.}(2023)]%
        {yu2023llm}
\bibfield{author}{\bibinfo{person}{Shengcheng Yu}, \bibinfo{person}{Chunrong Fang}, \bibinfo{person}{Yuchen Ling}, \bibinfo{person}{Chentian Wu}, {and} \bibinfo{person}{Zhenyu Chen}.} \bibinfo{year}{2023}\natexlab{}.
\newblock \showarticletitle{Llm for test script generation and migration: Challenges, capabilities, and opportunities}. In \bibinfo{booktitle}{\emph{2023 IEEE 23rd International Conference on Software Quality, Reliability, and Security (QRS)}}. IEEE, \bibinfo{pages}{206--217}.
\newblock


\bibitem[Zhang et~al\mbox{.}(2023)]%
        {zhang2023survey}
\bibfield{author}{\bibinfo{person}{Quanjun Zhang}, \bibinfo{person}{Chunrong Fang}, \bibinfo{person}{Yang Xie}, \bibinfo{person}{Yaxin Zhang}, \bibinfo{person}{Yun Yang}, \bibinfo{person}{Weisong Sun}, \bibinfo{person}{Shengcheng Yu}, {and} \bibinfo{person}{Zhenyu Chen}.} \bibinfo{year}{2023}\natexlab{}.
\newblock \showarticletitle{A survey on large language models for software engineering}.
\newblock \bibinfo{journal}{\emph{arXiv preprint arXiv:2312.15223}} (\bibinfo{year}{2023}).
\newblock


\end{thebibliography}

\balance

\end{document}